\documentclass[aps,prl,twocolumn,showpacs]{revtex4}
\usepackage{epsfig}
\usepackage{graphicx}

\begin{document}

\DeclareGraphicsExtensions{.eps,.EPS}

\title{Non-equilibrium quantum magnetism in a dipolar lattice gas}

\author{A. de Paz$^{1,2}$, A. Sharma$^{2,1}$, A. Chotia$^{2,1}$, E. Mar\'echal$^{2,1}$,  J. H. Huckans$^{4,1}$, P. Pedri$^{1,2}$, L. Santos$^3$, O. Gorceix$^{1,2}$, L. Vernac$^{1,2}$ and B. Laburthe-Tolra$^{2,1}$}

\affiliation{$^{1}$\,Universit\'e Paris 13, Sorbonne Paris Cit\'e, Laboratoire de Physique des Lasers, F-93430, Villetaneuse, France\\
$^{2}$\,CNRS, UMR 7538, LPL, F-93430, Villetaneuse, France\\
$^{3}$\, Institut f\"ur Theoretische Physik, Leibniz Universit\"at Hannover, Appelstr. 2, DE-30167 Hannover, Germany\\
$^{4}$\, Department of Physics and Engineering Technology, Bloomsburg University of Pennsylvania, Bloomsburg, PA 17815, USA}

\begin{abstract}

Research on quantum magnetism with ultra-cold gases in optical lattices is expected to open fascinating perspectives for the understanding of fundamental
problems in condensed-matter physics. Here we report on the first realization of quantum magnetism using a degenerate dipolar gas in an optical lattice.
In contrast to their non-dipolar counterparts, dipolar lattice gases allow for inter-site spin-spin interactions without relying on super-exchange energies, which
constitutes a great advantage for the study of spin lattice models. In this paper we show that a chromium gas in a 3D lattice
realizes a lattice model resembling the celebrated $t$-$J$ model, which is characterized by a non-equilibrium spinor dynamics resulting
from inter-site Heisenberg-like spin-spin interactions provided by non-local dipole-dipole interactions.
Moreover, due to its large spin, chromium lattice gases constitute an excellent environment for the study of quantum magnetism of high-spin systems, as
illustrated by the complex spin dynamics observed for doubly-occupied sites.

\end{abstract}

\date{\today}
\maketitle



The study of quantum magnetism is of utmost importance for the understanding of a variety of modern materials with strong correlations \cite{Auerbach}. Cold atoms loaded in periodic optical potentials provide a new platform for the investigation of quantum magnetism that presents several important interesting features, such as the absence of unwanted disorder, and the possibility to tune the inter-particle interactions \cite{RevBloch}. This results in a well defined Hamiltonian correctly describing the system. For this reason, there has been in recent years a huge interest towards using atoms in optical lattices as quantum simulators
for various theoretically intractable many-body problems \cite{simulator}, relevant for phenomena such as high-$T_c$  superconductivity.



\begin{figure*}
\centering
\includegraphics[width=2\columnwidth]{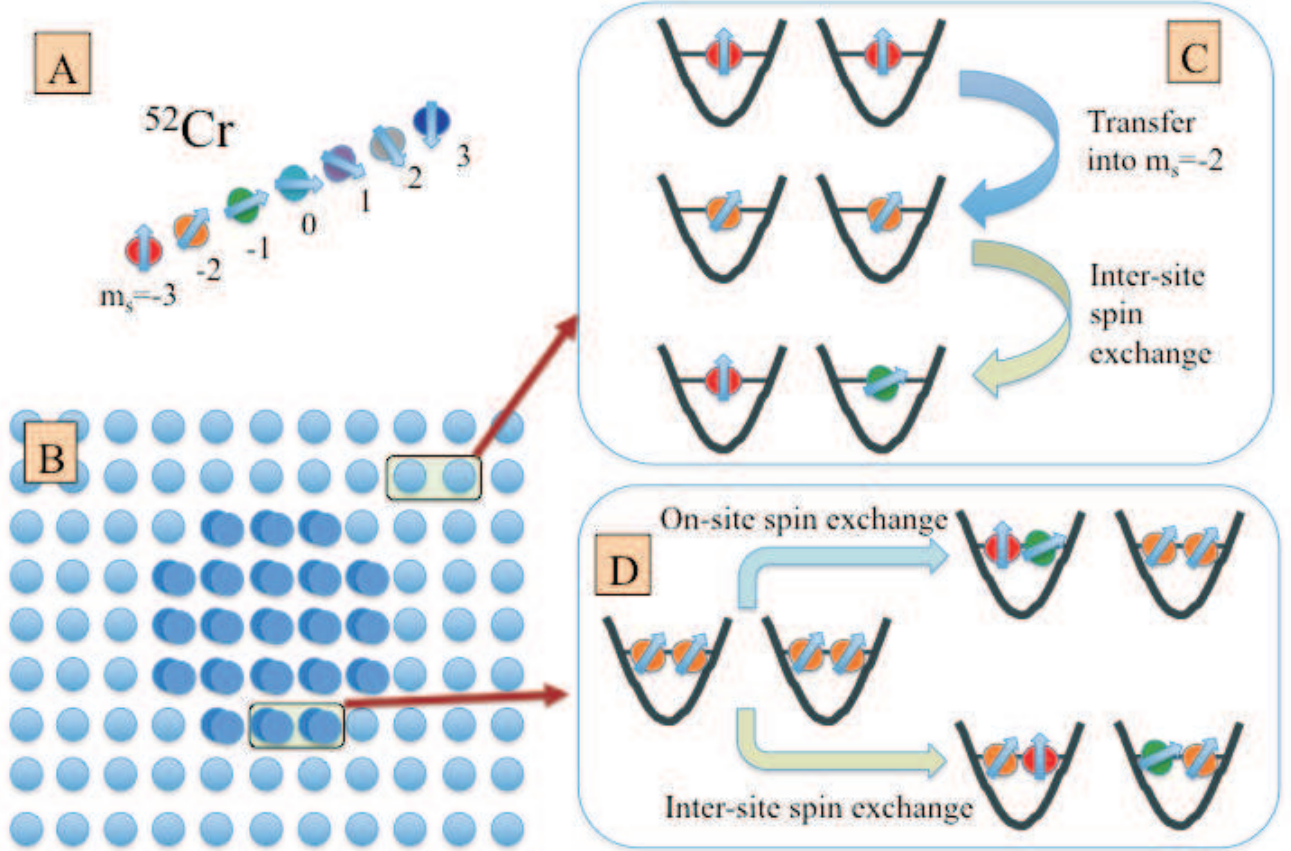}
\caption{\setlength{\baselineskip}{6pt} {\protect\scriptsize \textbf{Principles of Non-equilibrium quantum magnetism in a dipolar lattice gas.}
(A) Bosonic chromium, $^{52}$Cr, possesses a large spin $s=3$, and hence presents $7$ possible Zeeman substates, $m_s=-3,\dots 3$. In our experiments, only $m_s=-3$, $-2$, $-1$ and
$0$ play a relevant role. (B) As indicated in the text, we load an initial BEC into a deep 3D lattice in the Mott regime. Due to the overall harmonic confinement, a central core is formed with
doubly occupied sites, surrounded by singly-occupied sites. (C) The atoms are initially prepared in $m_s=-3$. Since spin relaxation cannot occur due to the external magnetic field
applied, no spin dynamics may occur if all atoms are in this state. We initiate the spin dynamics by transferring the atoms into $m_s=-2$. Singly-occupied sites may undergo
spin-exchange only through inter-site dipole-dipole interactions. (D) In contrast, doubly occupied sites may present both on-site and inter-site spin-exchange. As shown in the text,
on-site exchange is responsible for fast short-time~($\sim 200$ $\mu$s) spin oscillations, whereas inter-site spin-exchange between doubly-occupied sites results in slow spin oscillations at a longer time scale~($\sim 3$ ms).}} \label{Principle}
\end{figure*}


Dilute atomic gases possess specific properties which result in qualitatively novel physics compared to that of electrons in solids. In particular, atoms may
carry a larger spin $s>1/2$ provided by their internal structure, which results in an exceedingly rich phenomenology~\cite{Review-Ueda,ReviewDSK}. In particular, atomic gases in optical lattices allow for the realization of quantum magnetism with $s>1/2$, which results in a wealth of novel quantum phases, as
spin nematics~\cite{LatticeDemler}, color superfluidity~\cite{SUN2}, or chiral spin liquids~\cite{SUN1}.

Dipolar gases, which may be realized either with magnetic atoms \cite{CrBEC,CrBECus,DysprosiumBEC,DysprosiumFermiSea,Erbium} or with polar molecules~\cite{Ni08,Chotia12,Wu12,Heo12},
open, due to the peculiar long-range anisotropic nature of the dipole-dipole interactions~(DDIs),
 fascinating possibilities for the engineering of models of quantum magnetism~\cite{QMH1,QMH2,QMH3,QMH4,Carr,RevueDipQ3}.
In non-dipolar gases, contact interactions do not couple atoms belonging to different lattice sites, and hence an effective inter-site
spin exchange can only result from tunneling-assisted super-exchange processes. As a result, effective spin-spin interactions are typically very weak~($\sim 1$ Hz)\cite{Trotzky}.
In contrast, the long range DDIs provide a direct inter-site spin-spin coupling \cite{BECinterfero,CrBECStability} which hence becomes independent of tunneling.
Moreover, inter-site spin coupling becomes significantly larger ($\sim 20$ Hz for the case of our chromium experiments) easing the constraints to study quantum magnetism
within the time scale set by the coherence time.

This paper pioneers, to the best of our knowledge, the experimental realization of quantum magnetism with a degenerate dipolar gas in an optical lattice.
We study the particular case of $^{52}$Cr bosons, a system that possesses not only unusually strong magnetic DDIs, but a large spin ($s=3$) as well, and hence
allows for the investigation of the rich physics expected for high-spin lattice gases.
Our experimental results on spin dynamics reveal for the first time spin exchange between different lattice sites mediated by the DDIs.
We first explore the case with maximally one atom per site, where spinor dynamics is purely provided by inter-site DDIs, and may be well described by
a model resembling the well known $t$-$J$ hamiltonian, a fundamental model of quantum magnetism~\cite{Auerbach}.
We then investigate the scenario where two atoms may occupy the same site, in which the interplay between contact and dipolar interactions lead
to a rich spin exchange physics characterized by two markedly different coherent spin oscillations.
Whereas on-site contact interactions lead to fast spin oscillations at short-time scales, coherent oscillations  of a much lower frequency are observed
at longer times. We show that the latter disappear when a magnetic field gradient is applied, hence demonstrating that
slow coherent oscillations result from DDIs between doubly occupied sites.



The DDI between two atoms having a dimensionless spin ${\bf S}_i$ is described by the Hamiltonian:
\begin{equation}
\hat{H_d}= d^2 \frac{{\hat{\bf S}_1}\cdot {\hat{\bf S}_2}-3({\hat{\bf S}_1}\cdot{\bf \hat{r}})({\hat{\bf S}_2}\cdot {\bf\hat{r}})}{4\pi r^3},
\label{eq:Hd}
\end{equation}
where $d^2=\mu_0(g \mu_B)^2$ ($\mu_0$ being the magnetic permeability of vacuum, $g$ the Lande factor, $\mu_B$ the Bohr magneton), and $\hat{\bf r}=\bf{r}$$/r$ with ${\bf r}={\bf r}_1-{\bf r}_2$ the interatomic separation.
The Hamiltonian~(\ref{eq:Hd})  includes magnetization changing terms, which may crucially modify the spinor physics by introducing free magnetization \cite{demagnetization}, and an intrinsic spin-orbit coupling \cite{EdH,Crtube}. However we have recently shown that these terms have a resonant character in a 3D lattice, and therefore are strongly suppressed at a low enough magnetic field $\bf{B}$~\cite{resonances}. In this case, considered throughout this paper, the spin projection along the $\bf{B}$ field is conserved, and
hence only Ising and exchange terms play a role. As a result, the hamiltonian of eq.(\ref{eq:Hd}) can be reduced to an affective hamiltonian \cite{Giovanazzi}, with a Heisenberg-like form~\cite{Auerbach}:
\begin{equation}
\!\! \hat{H}_d^{eff}\! =\!  \frac{d^2}{4\pi r^3}\! \left(1-3 \frac{z^2}{r^2} \right)\! \left(\hat{S}_{1z}\hat{S}_{2z} \! - \! \frac{1}{4}\left(\hat{S}_1 ^+ \hat{S}_2 ^- \! + \!  \hat{S}_1 ^- \hat{S}_2 ^+\right)
\! \right)
\label{HdSE}
\end{equation}
with $z$ the relative coordinate along the $\bf{B}$ field. The experiments detailed below reveal the spinor dynamics induced by inter-site spin-spin interactions of this
Heisenberg-like form, showing that a chromium gas loaded in a 3D lattice provides an interesting platform for the analysis of quantum magnetism.



In our experiment, we first create a chromium Bose-Einstein condensate~(BEC) in a crossed-beam optical dipole trap as described in Ref.~\cite{CrBECus}. The condensate, comprising about $10^4$ atoms polarized in the absolute ground Zeeman state $m_s=-3$, is confined at the bottom of a harmonic trap with frequencies  ($\omega_x$, $\omega_y$, $\omega_z$) = $2 \pi \times (400, 550, 300)$ Hz, within a magnetic field of $B=10$ mG. The BEC is loaded adiabatically in a 3D optical lattice, generated using $4$ W of a single-mode laser with wavelength $\lambda=532$ nm. The 5 beams architecture of our anisotropic lattice is described in ~\cite{resonances}. It consists of a rectangular lattice of periodicity $\lambda/2\times(1,1/\sin(\pi/8),1/\cos(\pi/8))$ along $x,y,z$ directions. The lattice depths in each direction are calibrated using Kapitza-Dirac diffraction~\cite{Kapitza} for pairs of beams, leading to a maximum of $30$ $E_R$ (with $E_R= h^2/2 m \lambda^2$ the recoil energy). In this paper (unless stated otherwise) we work at such lattice depth, which corresponds to lattice band gaps $(\omega_x^L$, $\omega_y^L$, $\omega_z^L$) = $2 \pi \times (100, 55, 170)$ kHz. For our experimental parameters, the predicted ground state is then a Mott state made of a central region with double occupancy surrounded by a single occupancy shell. The superfluid to Mott-insulator transition is reached at about $12 E_R$ \cite{Greiner}. Although the lattice geometry is complicated by its anisotropy \cite{resonances}, and we do not expect a fully adiabatic crossing towards the Mott state, the $30$ ms loading time allows to remain close to the ground state. This is confirmed by adiabatically ramping down the lattice at the same speed, in which case a BEC with only few excitations is recovered. The lattice band gaps are much larger than any other energy scales in the system (interactions, temperature, Zeeman shifts), and hence the atoms remain confined in the lowest energy band of the lattice.



\begin{figure}[!h]
\centering
\includegraphics[width=3.2in]{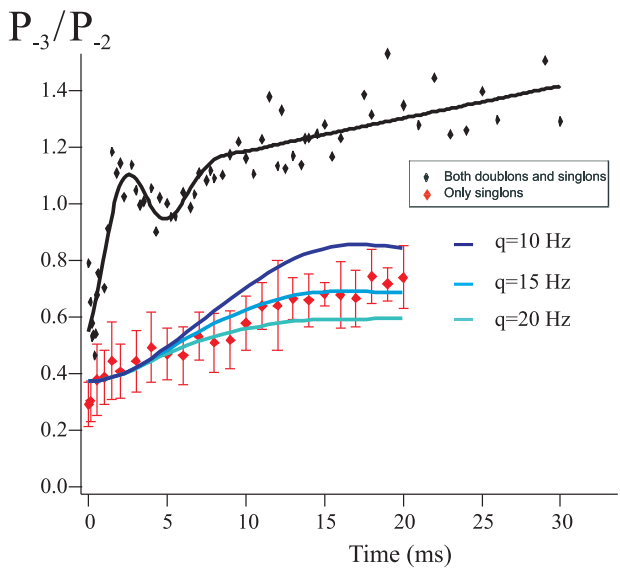}
\caption{\setlength{\baselineskip}{6pt} {\protect\scriptsize \textbf{Long term spin dynamics.} The ratio between two Zeeman spin components is plotted as a function of time. For this specific data, the lattice depth corresponds to lattice band gaps of $(\omega_x^L$, $\omega_y^L$, $\omega_z^L$) = $2 \pi \times (130, 55, 150)$ kHz. Experimental results are shown for singlons only (red diamonds) and for singlons plus doublons (black diamonds). Error bars show statistical uncertainties. The two dynamics show the same trend after 10 ms. The black line is a guide for the eye. The colored line are results of our simulation for spins 3 on a $3\times3$  plaquette, for three different quadratic effects, and tunneling rates $J_{(x,z)}=(11,3)$ Hz (deduced from lattice calibrations).}} \label{SinglonsVsDoublons}
\end{figure}




To demonstrate the existence of inter-site spin exchange, we selectively remove atoms which share a same site by inducing dipolar relaxation at high B field ($>300$ mG). Atoms are first transferred to the highest energy Zeeman state, which induces losses selectively for doubly occupied sites as the released energy is larger than the lattice depth \cite{resonances} (see $Methods$). We then transfer the remaining atoms, all belonging to sites with single occupation~(singlons), to the lowest energy Zeeman state ($m_s=-3$) and lower the B field to $10$ mG. After this preparation the system consists of a shell of about 4000 singly occupied sites close to unit filling. We initiate spin dynamics by transferring atoms to $m_s=-2$ (see Supplementary Material S1), with an efficiency of up to 80$\%$. We then let the spin populations evolve and make a Stern-Gerlach analysis after a given time t to measure the different spin populations.

Populations in states $m_s=-3,-2,-1,0$ evolve as a function of time in a typical time scale of 10 ms while the magnetization remains constant within error bars. To maximize signal to noise ratio, we plot the ratio of populations in $m_s=-3$ and $m_s=-2$ as a function of t in Fig. \ref{SinglonsVsDoublons}. Given that no site contains more than one atom, the observed spin dynamics is necessarily the product of an inter-site spin exchange process. In particular, we stress that at our lattice depth tunneling producing double occupancy is strongly suppressed by the on site contact interactions~($\simeq 10$ kHz) and any associated spin exchange process has a rate below $0.1$ Hz (set by superexchange interactions \cite{Trotzky}), incompatible with our observations. The spin dynamics shown in Fig. \ref{SinglonsVsDoublons} is therefore a direct demonstration of
dipolar inter-site spin exchange.

The spin dynamics of the singlon gas may be well understood from a 2D $t$-$J$-like Hamiltonian~\cite{Auerbach}.  The 2D assumption is well justified due to the anisotropy of the lattice spacing,
which is basically equal along $x$ and $z$ but approximately $3$ times larger along $y$. The $t$-$J$-like Hamiltonian acquires the form:
\begin{eqnarray}
\hat H&=& - \sum_{\langle i,j \rangle, m_s } J_{i,j}\hat b_{i,m_s}^\dag \hat b_{j,m_s} +q\sum_{j,m_s} m_s^2 \hat n_{j,m_s} \nonumber \\
&+& \sum_{i,j} V_{i,j} \left (\hat S_i^z\hat S_j^z-\frac{1}{4}\left ( \hat S_i^+ \hat S_j^- + \hat S_i^- \hat S_j^+ \right ) \right )
\label{eq:Hamiltonian}
\end{eqnarray}
Here $\langle \dots \rangle$ denotes nearest neighbors,  $\hat b_j$ and $\hat b_j^\dag$ the bosonic
annihilation and creation operators of particles at site $j=(j_x,j_z)$ with spin projection $m_s$,
$\hat n_{j,m_s}=\hat b_{j,m_s}^\dag \hat b_{j,m_s}$, and $\hat n_j=\sum_{m_s} \hat n_{j,m_s}$.
The hard-core constraint, $n_j=0,1$, is justified by the large on-site contact interaction associated to doubly-occupied sites.
Hopping, characterized by the rate $J_{i,j}$, is however possible due to the presence of residual empty sites. Estimated tunneling rates along the $(x,z)$ directions are (11,3) Hz~\cite{Zwerger02}.
We include as well the effect of a residual quadratic light shift, which introduces a spin-dependent energy for the singly occupied sites $q m_s^2$. This shift
may significantly handicap inter-site spin-changing collisions. Interferometric measurements~(see S4) provide an upper value of $25$ Hz for $q/h$.
Singly-occupied sites interact with each other through the Heisenberg-like interaction~(\ref{HdSE}), where the coupling constants $V_{ij}$ are evaluated
taking into account the spatial extension of the on-site wave function, as discussed in the Supplementary Material~(S2).
We have analyzed the exact quantum dynamics of the many-body system for a $3\times 3$ plaquette
using exact diagonalization employing periodic boundary conditions. In order to evaluate the effect of the motion of residual holes left behind in the preparation process,
we consider among the singly occupied sites one hole, initially localized at one site. We find that hopping, although not fully negligible, does not play a major role in our experiment, due to the rather strong lattice employed. As shown in Fig. \ref{SinglonsVsDoublons} the results of the plaquette calculation for $q/h\simeq 15$ Hz are in very good agreement with our experimental data, confirming that the
observed spin dynamics results from the Heisenberg-like interactions in eq.(\ref{eq:Hamiltonian}).




\begin{figure}[!h]
\centering
\includegraphics[width=3.2in]{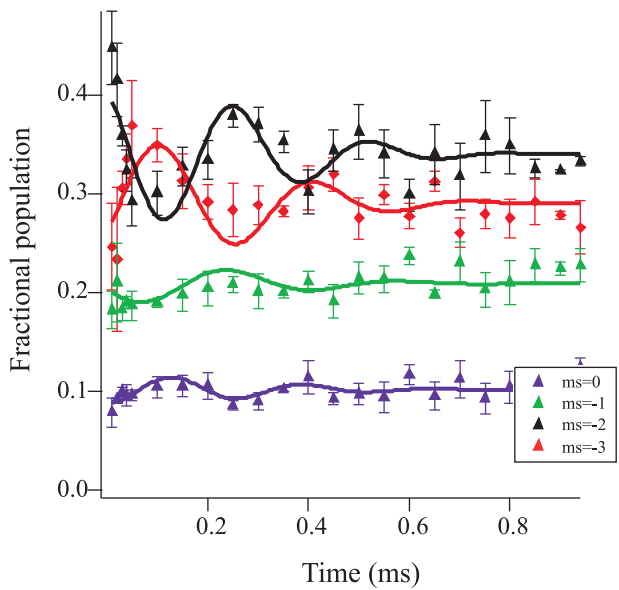}
\caption{\setlength{\baselineskip}{6pt} {\protect\scriptsize \textbf{Fast spin exchange dynamics due to contact interactions.} The experimental evolution of the different spin components shows damped oscillations. For this specific data, the lattice depth was reduced, corresponding to lattice band gaps of $(\omega_x^L$, $\omega_y^L$, $\omega_z^L$) = $2 \pi \times (100, 50, 145)$ kHz. The value of the pseudo period for $m_s=-3,-2$ is in good agreement with theory. Full lines are results of fit with damped exponentials.}} \label{Contact}
\end{figure}



\begin{figure}[!h]
\centering
\includegraphics[width=3.2in]{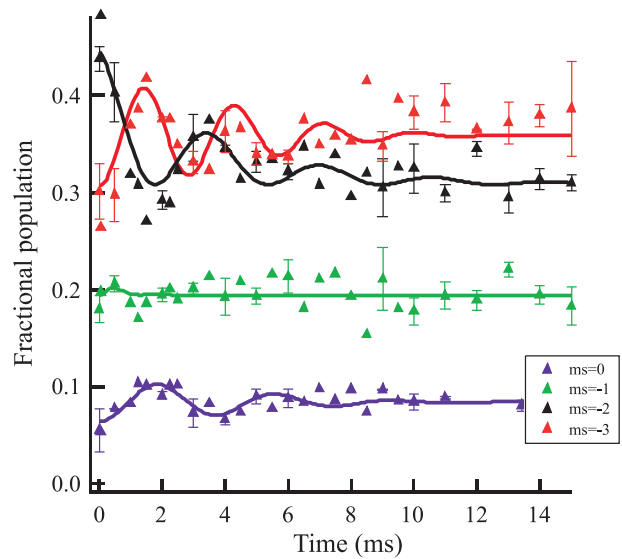}
\caption{\setlength{\baselineskip}{6pt} {\protect\scriptsize \textbf{Slow spin exchange dynamics due to dipole-dipole interactions.} The experimental evolution of different spin components shows oscillations on a few ms time scale. Lines are results of fits to the data with damped exponentials.}} \label{IntersiteDoublons}
\end{figure}




We now discuss the case where we do not empty doubly occupied sites (doublons). As shown below this scenario is characterized
by the rich interplay between contact interactions and DDI in the $s=3$ spin gas.
The system consists of a core of doublons surrounded by a shell of singlons, following the "wedding cake" atomic distribution characteristic of a trapped Mott state \cite{weddingcake}. We operate also in this case at a B field ($\simeq10$ mG) much larger than the critical field for spontaneous demagnetization \cite{demagnetization} that precludes any spin dynamics when atoms are prepared in $m_s=-3$. The state preparation at t=0 is identical to the singlon case, and the subsequent spin dynamics is illustrated in Figs. \ref{SinglonsVsDoublons}, \ref{Contact}, and \ref{IntersiteDoublons}) . At a short time scale, we observe fast spin oscillations~(Fig.\ref{Contact}), which damp in a few hundred $\mu$s. Then a second and slower dynamics occurs~(Fig. \ref{IntersiteDoublons}), showing two to three spin oscillations with a period of about 3 ms. After these oscillations are damped, we observe a slow drift of populations similar to the case of singlons, as shown in Fig. \ref{SinglonsVsDoublons}. All the observed spin dynamics occurs at constant magnetization.


The fast spin oscillations shown in Fig. \ref{Contact} result from on-site spin dependent contact interactions.
For an isolated site with two atoms, the eigenstates are characterized by a total spin $S$ and spin projection $M$ along the $B$ field.
A state of two atoms in $m_s=-2$ in the same lattice site constitutes a linear superposition of states with $M=-4$ and $S=6$ and $4$,
$\sqrt{\frac{6}{11}}\left|6,-4\right>-\sqrt{\frac{5}{11}}\left|4,-4\right>$.  Thus a Rabi-like oscillation \cite{Widera} is expected, with a period corresponding to the beating of the
two eigenfrequencies:
\begin{equation}
T_c=\frac{h}{n_0|g_6-g_4|}
\label{PeriodSEcontact}
\end{equation}
where $n_0=\int d^3r |\psi_0({\bf r})|^4$, with $\psi_0({\bf r})$ the on-site wave function, and $g_S=4 \pi \frac{\hbar^2}{m} a_S$ \cite{Stringari}, with $m$ the mass of the atoms and $a_S$ the $s$-wave scattering length associated with the scattering channel with total spin $S$. For the data in Fig. \ref{Contact}, $n_0\simeq 6.3$ $10^{20}$ m$^{-3}$.
Employing the values of $a_6$ and $a_4$ from Ref.~\cite{PRA RD}, we then obtain a theoretical oscillation period $T_c\simeq 320\pm(50)$ $\mu$s, in good agreement with the period of the oscillations of the populations in $m_s=-3,-2$ shown in Fig. \ref{Contact}, $280\pm(30)$ $\mu$s. These results constitute the first observation of spin exchange dynamics due to contact interactions in a $s=3$ spinor gas. Note that on-site spinor dynamics is much faster in Chromium than in other spinor gases~\cite{Widera, Krauser},
due to the much larger difference between the scattering lengths of the relevant collision channels. We also observe a much stronger damping (of about $2$ kHz) than in ~\cite{Widera, Krauser}. The full analysis of spin exchange oscillations due to contact interactions will be the subject of a forthcoming paper.

\begin{figure}[!h]
\centering
\includegraphics[width=3.2in]{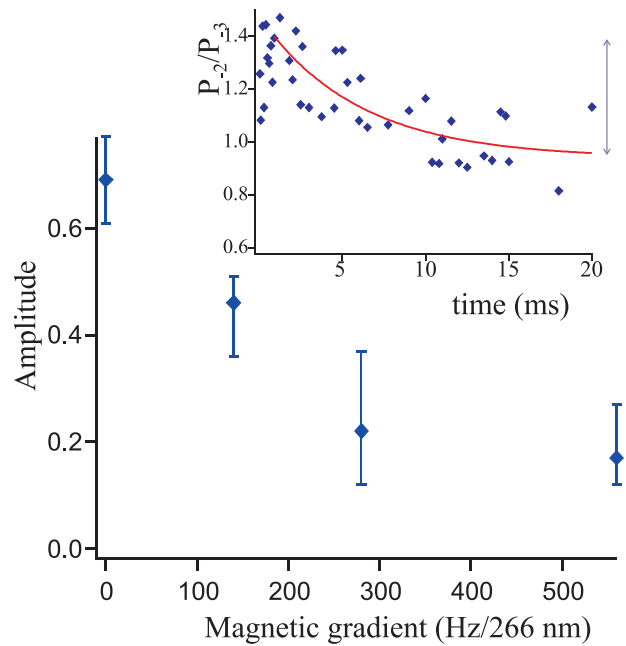}
\caption{\setlength{\baselineskip}{6pt} {\protect\scriptsize \textbf{Effect of a magnetic gradient on the amplitude of the slow spin dynamics.} The inset shows fit of the data with an exponential from which we deduce the amplitude (indicated by the arrow) plotted in the main figure.}} \label{Gradient}
\end{figure}

At longer time scales (see Fig. \ref{IntersiteDoublons})
the observed spin oscillations present a characteristic frequency much smaller than that of the short-time oscillations.
To demonstrate that these ms-scale oscillations and in general the long-time spin dynamics result mostly from inter-site DDIs, we have applied magnetic field gradients, $\Delta B$, to our sample. These gradients induce Zeeman energy shifts between adjacent sites. When these shifts are larger than the non local interaction, inter-site spin dynamics is energetically forbidden. We apply a few G.cm$^{-1}$ (note that $1$ G.cm$^{-1}$ corresponds to a Zeeman shift of $\Delta E =70$ Hz between two adjacent sites in our lattice). The magnetic field remains below the first resonance for dipolar relaxation across the entire sample. The energy shifts associated to the magnetic field gradient lie also below the excitation gap for the Mott state \cite{Greiner}. The effect of the magnetic gradient on the amplitude of the spin dynamics is shown in Fig. \ref{Gradient}. We derive the amplitude of the dynamics by fitting the ratio between the spin populations in $-2$ and $-3$ with an exponential, as shown in the inset. We observe a strong suppression of the spin dynamics amplitude in presence of gradients, which shows that the spin dynamics involves a non local coupling between atoms. The value of $\Delta B$ at which the amplitudes significantly drop corresponds to $\Delta E/h$
of the same order as the frequency of the oscillations shown in Fig. \ref{IntersiteDoublons}. The small residual spin dynamics at large gradients can be attributed to on site interactions.

Hence, spin dynamics at t$>1$ ms result mostly from inter-site DDIs. Moreover, as shown in Fig. \ref{SinglonsVsDoublons}, spin oscillations at the ms timescale disappear when doublons are removed. These observations provide a strong evidence that the long-time spin oscillations are due to intersite DDIs between doubly-occupied sites. Note that, interestingly,
doubly-occupied sites lead to stronger DDIs compared to single $^{52}$Cr atoms, as pairs of atoms behave like molecular Cr$_2$ magnets with larger magnetic moments, without the actual need of creating the molecules using Feshbach resonances.



To qualitatively account for the role of the DDIs between doublons in the long-time spin oscillations, we have developed a toy model consisting of two doubly-occupied sites, for which
$12$ possible states may be dynamically reached via inter-site DDIs for a preserved overall magnetization $M_t=-8$. We show in S3 that the total spin $S$ of a pair of particles in one site is not modified by the interaction with other sites, which confirms that pairs of particles do behave like large spins $S$ interacting through long range DDIs. Starting with an initial state $\left| S=6,M=-4\right>$ in both sites, our model qualitatively reproduces the shape of the oscillations shown in Fig. \ref{IntersiteDoublons}: $m_s=-3$ and $m_s=-2$ oscillate out of phase, $m_s=-1$ shows almost no oscillation, and $m_s=0$ oscillate in phase with $m_s=-3$ (see S3). Nevertheless this toy model cannot reproduce the observed time scale for the oscillations, as it only includes the interactions between two sites, while each site is in fact coupled to many (typically 6) sites due to the long range character of DDIs. Our experiment shows that the frequency of the spin oscillations is about 7 times faster than the one predicted by the two site model. This indicates that a much more elaborate many-body theoretical treatment would be necessary to interpret this intriguing spin dynamics, with an interplay between short range physics and long range interactions between many sites. We note that our system may present similarities with the physics of Frenkel excitons, which could be studied with polar molecules, for which similar damped oscillations are predicted \cite{Krems}.


In conclusion, we have demonstrated for the first time inter-site spin-exchange in a dipolar gas due to the long-range character of the dipole-dipole interactions. Dipole-induced spin-exchange, being much larger than that resulting from super-exchange in non-dipolar gases in deep optical lattices, opens fascinating perspectives for the study of quantum magnetism with magnetic atoms. We have in particular shown that a chromium lattice gas with maximally one atom per site presents purely inter-site spin exchange, realizing a lattice model resembling the $t$-$J$ model of strongly-correlated electrons. We have studied in addition the involved spinor dynamics resulting from doubly-occupied sites, which stems from the interplay between
spin-dependent contact and dipole-dipole interactions. We have shown in particular that fast short-time spin oscillations result from spin-changing contact interactions, constituting the first observation of these interactions in a $s=3$ gas.
In contrast, we have demonstrated that longer-time coherent spin oscillations are the result of inter-site spin-exchange between doubly-occupied sites. Our experiment demonstrates hence not only the potential of dipolar gases for the quantum simulation of lattice models, but also the intriguing physics associated to lattice gases of large spin.

Acknowledgements:
We acknowledge financial support from Conseil R\'{e}gional d'Ile-de-France and from Minist\`{e}re de l'Enseignement Sup\'{e}rieur et de la Recherche within IFRAF and CPER. L. S. acknowledges support from the Deutsche Forschungsgemeinschaft (SA1031/6) and the Cluster of Excellence QUEST.

\vspace{0.5cm}

\centerline{\Large \textbf{Methods}}

\paragraph{Suppression of multiply-occupied sites}

In order to work with sites occupied by a single atom, we use dipolar relaxation at a B field large enough such that these inelastic collisions release enough energy to expel the atoms out of the lattice. When the Larmor frequency is larger than $V_0/\hbar$ ($V_0$ being the lattice depth), we observe a non-exponential loss dynamics: the atom number rapidly decreases in 10 ms before reaching a non-zero steady value and then no loss are observed for tens of ms. Losses are due to dipolar relaxation between atoms occupying the same site and they stop when only single occupancy sites remain.

\vspace{0.5cm}
\centerline{\textbf{Note added}}
After the completion of this manuscript, we have become aware of a very recent preprint by the group of D. Jin and J. Ye. This group reports on the observation of intersite DDIs between pairs of dipolar heteronuclear molecules trapped in an optical lattice, at low filling factors \cite{Ye}.

\newpage
\null
\centerline{\Large \textbf{Supplemental material}}

\vspace{0.5cm}

\paragraph{S1.Atomic state preparation}

\subparagraph{Principle}
To initiate the spin dynamics in our experiment, we transfer the atoms from $m_s=-3$ to $m_s=-2$ through an adiabatic Raman transfer. This transfer is performed by ramping up a quadratic (i.e. tensor) light shift which leads to a level crossing between states $m_s=-3$ and $m_s=-2$, above which $m_s=-2$ is the Zeeman ground state. A weak two photon coupling between $m_s=-3$ and $m_s=-2$ ensures the transfer.

When the Zeeman levels are equally spaced it is not possible to isolate a two-level system, and state preparation in a well defined Zeeman state is difficult (apart from fully stretched Zeeman states). As there is no hyperfine structure for $^{52}$Cr, we rely on quadratic light shifts (QLS, i.e. energy shifts equal to $q\times m_S^2$) arising from the tensorial nature of atom-light coupling, to break the linear Zeeman energy ordering. Significant QLS can be obtained with a laser beam close to an atomic transition. We use a 427.85 nm laser, red detuned from the $^7S_3\rightarrow ^7P_3$ transition at 427.6 nm. Then calculations show (see for example \cite{EPJD}) that the light shifts of the Zeeman states are almost purely quadratic (the linear part is almost suppressed) with a $\sigma^-$ polarized light. For a $1/e^2$ radius Gaussian beam of 100 $\mu$m, and a power of 40 mW, we estimate $q=10$ kHz. A degeneracy between states $m_s=-3$ and $m_s=-2$ is obtained for $q=8$ kHz at a Larmor frequency $2g_S\mu_B/h=40$ kHz.

An efficient transfer from $m_s=-3$ to $m_s=-2$ is provided by a two photon Raman transition, if the laser contains a small admixture of $\pi$ light. For example for an intensity percentage of $\pi$ light $p=0.3\%$ (corresponding to an angle between the laser beam propagation axis and the direction of the B field equal to $4^{\circ}$), we obtain a Raman frequency $\Omega_R=\sqrt{p}\frac{\gamma^2}{4\Delta}s_0 \left<3,-3;1,0\right|3,-3\left>\right<3,-2;1,-1\left|3,-3\right>=2\pi\times20$ kHz (with $\gamma=3.07\times10^7$s$^{-1}$ the linewidth of state $^7P_3$, $\Delta$ the detuning relative to $^7P_3$, $\left<j_1,m_{j1};j_2,m_{j2}\right|J,M_J\left>\right.$ are Clebsch -Gordan coefficients, and $s_0=\frac{I_L}{I_{sat}}$: $I_L$ is the maximal laser intensity, $I_{sat}=82$ W.m$^{-2}$).

In practice, we linearly ramp the laser power in 1.5 ms from zero to the maximal power, thus adiabatically transferring the atomic population to $m_s=-2$. We then abruptly reduce the power to zero (in about 1 $\mu$s) to produce an excited (Zeeman) state from which we observe spin dynamics.
The photon scattering rate from $m_s=-2$ is evaluated to 3.s$^{-1}$ for $\sigma^-$ polarization, so that photon scattering is negligible for ramps shorter than a few ms.

\begin{figure}[!h]
\centering
\includegraphics[width=3in]{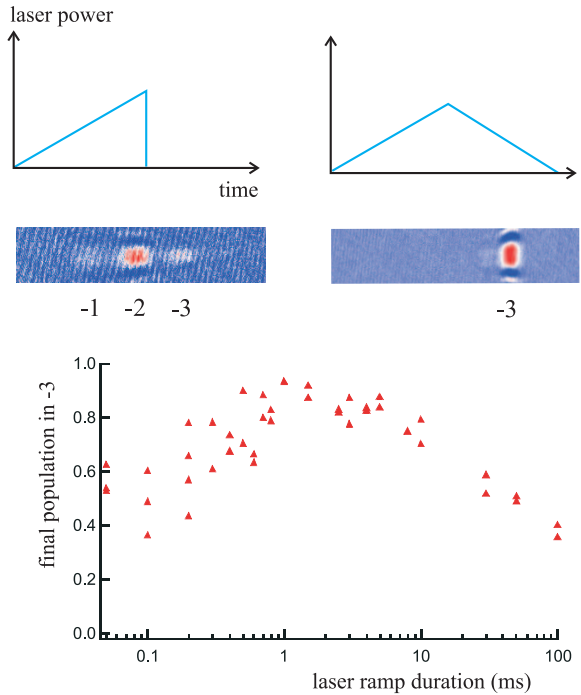}
\caption{\setlength{\baselineskip}{6pt} {\protect\scriptsize \textbf{Characterization of the adiabaticity of the Raman preparation}. We show absorption images of the atomic sample after Stern Gerlach separation of the Zeeman components following (left) a non symmetric (with fast return to zero) and (right) a symmetric ramp of the laser power. Provided an adequate duration for the ramp, we can obtain in the first case an excited atomic sample with most ($>80\%$) of the atoms transferred in $m_s=-2$ , while in the second case we recover a full polarized sample in $m_s=-3$. The bottom curve shows the final population in $m_s=-3$ after the symmetric ramp as a function of the ramp duration $t$: for $t<1$ ms, the ramp is too fast and the transfer in $m_s=-2$ is not adiabatic; for $t>10$ ms, photons can be scattered, which alters the final state. }} \label{AdiabRevers}
\end{figure}

\vspace{0.2cm}
\subparagraph{Performance and Reversibility}
We show in Fig. \ref{AdiabRevers} the influence of the ramp duration on the performance of the adiabatic transfer. The transfer to $m_s=-2$ is efficient provided that the ramp is longer than 1 ms, to ensure adiabaticity, and no longer than 10 ms, to prevent significant photon scattering. We recover the initial atomic population in $m_s=-3$ for a 1 to 5 ms long symmetric ramp, which proves adiabaticity.

\vspace{0.2cm}
\subparagraph{Effect of a transverse magnetic field}
\begin{figure}[!h]
\centering
\includegraphics[width=3.4in]{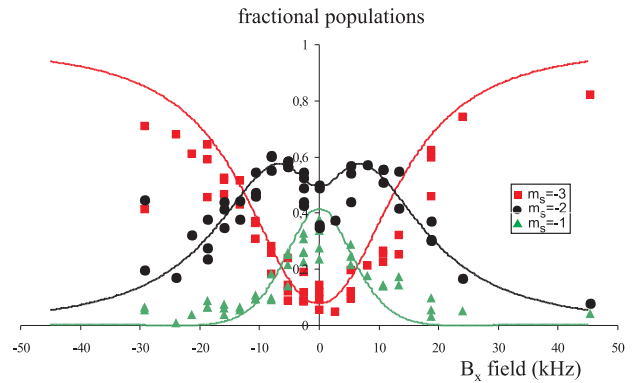}
\caption{\setlength{\baselineskip}{6pt} {\protect\scriptsize \textbf{Test of the validity of the Raman model}. We plot the final population in $m_s=-1,-2,-3$ as a function of a transverse component $B_x$ of the magnetic field, for a given value of the longitudinal component ($B_z$=15 kHz). Points: experimental results. Full lines: results of the calculations of our Raman model, assuming $B_y=5$ kHz.
 }}
\label{AngleBLaser}
\end{figure}
In order to confirm the validity of the Raman model to interpret the state preparation, we have studied the influence of the angle $\theta$ between the laser beam and the \textbf{B} field. We plot on Fig. \ref{AngleBLaser} the fractional population transferred in $m_s=-3,-2,-1$ at the end of the ramp, as a function of a B field component orthogonal to the laser $B_x$. We kept the longitudinal component of the \textbf{B} field $B_z$ constant (with a corresponding Larmor frequency of 15 kHz). The laser power and wavelength were respectively 45 mW and 427.9 nm; the linear laser power ramp was 1.5 ms.  We find a good agreement between experiment results and our calculations provided a constant non zero transverse component along the $y$ axis, $B_y=5$ kHz. For the experimental parameters of this study, the Raman ramp intensity is strong enough that some population is transferred as well to $m_s=-1$.

\vspace{0.5cm}

\paragraph{S2.Description of the plaquette simulation for singlons}

As mentioned in the main text, the spinor dynamics for the case of maximally one atom per site~(singlon gas) may be described using a $t$-$J$-like model.
In this section, we discuss in more details some relevant issues concerning this model and its numerical simulation.

\begin{figure}[!h]
\centering
\includegraphics[width=3in]{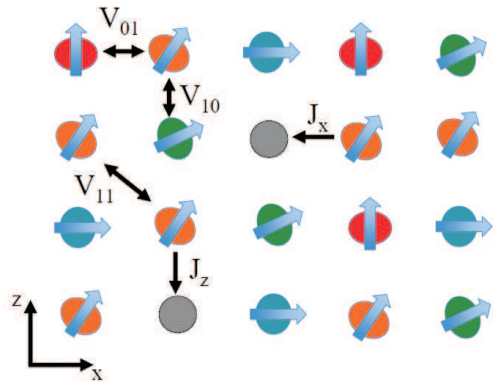}
\caption{\setlength{\baselineskip}{6pt} {\protect\scriptsize \textbf{Schematics of the $t-J$ model.} We show in this picture the different parameters included in our plaquette model. Our system realizes in practice an anisotropic $t-J$ model.}} \label{Plaquette}
\end{figure}

The coupling constants $V_{ij}$ between two sites $i$ and $j$ in Eq.~(3) of the paper depend on the vector separating both sites,
$\frac{\lambda}{2}{\mathbf n}$, with ${\bf n}=(n_z,n_x)$, and may be evaluated
assuming a Gaussian dependence of the on-site wave function
\begin{equation}
\psi_0({\bf r})=\frac{1}{\pi^{3/4}\sqrt{l_xl_yl_z}} e^{-x^2/2l_x^2}e^{-y^2/2l_y^2}e^{-z^2/2l_z^2}
\label{eq:psi0}
\end{equation}
where $l_j=\sqrt{\hbar/m\omega_j^L}$, with $\omega_j^L$ the effective on-site oscillator frequency in the $j=x,y,z$ direction.
Working in momentum space, using convolution theorem and employing the Fourier transform
of the dipole-dipole interaction the coupling constant acquires the form
\begin{eqnarray}
V_{{\bf n}}&=& U_0
\int \frac{d^3q}{(2\pi)^3}\frac{4\pi}{3} \left ( \frac{3q_z^2}{q^2}-1 \right ) e^{i(q_x n_x+q_z n_z)} \nonumber \\
&\times& e^{-\frac{q_x^2L_x^2}{2}-\frac{q_y^2L_y^2}{2}-\frac{q_z^2L_z^2}{2}} ,
\label{eq:Vdelta}
\end{eqnarray}
with $L_j=l_j/(\lambda/2)$, and $U_0=\frac{\mu_0\mu_B^2}{\pi(\lambda/2)^3}\simeq 2.8$Hz in our experiment. For the lattice employed for the experiments depicted in
Fig.~2 of the paper we obtain $V_{(0,1)}/U_0=0.770$, $V_{(1,0)}/U_0=-1.823$, and $V_{(1,1)}/U_0=-0.114$. Due to periodic boundary conditions we just consider these
three interactions in our $3\times 3$ plaquette calculations, but in general these are of course the most relevant ones. Note that neglecting the spatial extension of the on-site wave functions would have resulted in
$1$, $-2$ and $-0.177$, respectively. The finite width of the on-site wave function hence leads to a significant direction-dependent reduction of the inter-site interaction.

We simulate the exact quantum dynamics given by the Hamiltonian of Eq.~(3) of the paper in a $3\times 3$ plaquette
by means of the fourth-order Runge-Kutta method employing periodic boundary conditions. Note that fluctuations play a crucial role in the spinor dynamics, precluding the use
of mean-field calculations based on e.g. a dynamical Gutzwiller Ansatz.
The main reason is the crucial role played by the inter-site Ising~($S^zS^z$) interactions in reducing the effective
spin-changing collision rate. This may be easily understood as follows. A spin-changing collision results in a transition between a given
initial many-body state and a final one. Although the two states just differ in the spin state at two sites, the energy resulting from Ising interactions with neighboring sites may be quite different.
This energy difference may be larger than the spin-exchange rate, and hence may handicap the spin-exchange.
Note that due to the non-local character of Ising interactions, the energy difference between many-body states linked by a spin exchange
crucially depends on the particular spin configuration of neighboring sites. This crucial effect is lost in mean-field-like treatments, based on a Gutzwiller ansatz
for the many-body state, $| \psi \rangle =\otimes_j |\psi_j\rangle$, with $|\psi_j\rangle =a_j |hole\rangle + \sum_{m_s} f_{m_s,j} |m_s\rangle$. We have checked that, as expected from the
previous discussion, a dynamical Gutzwiller ansatz calculation leads
to a much faster spinor dynamics as that observed in our experiments. The time scale of the spinor dynamics is on the contrary well recovered by the exact calculations,
as shown in Fig.~2 of the paper, demonstrating that the dynamics is crucially determined by inherently quantum many-body effects.

\vspace{0.5cm}

\paragraph{S3. Two-site model for the spin dynamics of doubly-occupied sites}

As mentioned in the paper, slow spin oscillations in a time scale of few ms are due to spin-exchange between doubly occupied sites.
In this section we discuss in more detail a simplified model describing these oscillations.

We consider two doubly-occupied sites. Isolated doubly-occupied sites are characterized by two-body molecular eigenstates, $| S,M\rangle$, with a given total spin $S$ and spin projection $M$
along the ${\bf B}$ field, with corresponding eigenenergies $E_{S,M}=2\mu_B B M+ g_S n_0 + V_{(0,0)} C_{S,M}$, where
${C_{S,M}}=\langle S,M  | \hat F(\hat{\bf S}_1,\hat{\bf S}_2) | S,M \rangle$, with
$\hat F(\hat{\bf S}_1,\hat{\bf S}_2)= \hat S_1^z \hat S_2^z-\frac{1}{4} ( \hat S_1^+ \hat S_2^- + \hat S_1^- \hat S_2^+)$ .
For the parameters of Fig.~4 of the paper, $V_{(0,0)}/h\simeq 52$ Hz. Considering states with $S=6$ and $S=4$,
the largest on-site dipolar shift, $470$ Hz, is experienced by $|6,-6\rangle$. Note hence that
$E_{S,M}$ is overwhelmingly determined by the on-site interactions~($\sim 10$ kHz).

Starting with molecular states $\left|6,-4\right>$ and $\left|4,-4\right>$, spin-exchange may result in
$12$ possible two-site states $|S_L,M_L\rangle \otimes |S_R,M_R\rangle$, which may be sub-divided into $4$ families characterized by the value of
$S_L=6,4$ and $S_R=6,4$ (see Table \ref{Table}). Interestingly,
inter-site DDI only couple states within each family, i.e. pairs of atoms at one site behave like true large $S=6$ or $S=4$
spins interacting through DDIs. The Hamiltonian matrix is hence block-diagonal of the form
$V_{I} \langle S_L,M_L; S_R,M_R | F(\hat{\bf S}_L,\hat{\bf S}_R) | S_L,M'_L; S_R,M'_R \rangle$.
The largest coupling, $-6.4\, V_{I}$, links the state $|6,-4\rangle \otimes |6,-4\rangle$ with $|6,-5\rangle \otimes |6,-3\rangle$.

\begin{table}[t]
\centering
\begin{tabular}{|c|c|c|c|}
\hline
$F_1$ & $F_2$ & $F_3$ & $F_4$\\
\hline
$\left|6,-2\right>$ $\left|6,-6\right>$&  &$\left|4,-2\right>$ $\left|6,-6\right>$  & \\
$\left|6,-3\right>$ $\left|6,-5\right>$ & &$\left|4,-3\right>$ $\left|6,-5\right>$ & \\
$\left|6,-4\right>$ $\left|6,-4\right>$ &$\left|6,-4\right>$ $\left|4,-4\right>$ & $\left|4,-4\right>$ $\left|6,-4\right>$& $\left|4,-4\right>$ $\left|4,-4\right>$\\
$\left|6,-5\right>$ $\left|6,-3\right>$ &$\left|6,-5\right>$ $\left|4,-3\right>$ & & \\
$\left|6,-6\right>$ $\left|6,-2\right>$ &$\left|6,-6\right>$ $\left|4,-2\right>$ & & \\
\hline
\end{tabular}
\caption{The 12 two-sites states forming 4 independent subspaces. $|S_L,M_L\rangle |S_R,M_R\rangle$ indicate the spin state in the left and right site.}
\label{Table}
\end{table}

\begin{figure}[!h]
\centering
\includegraphics[width=3in]{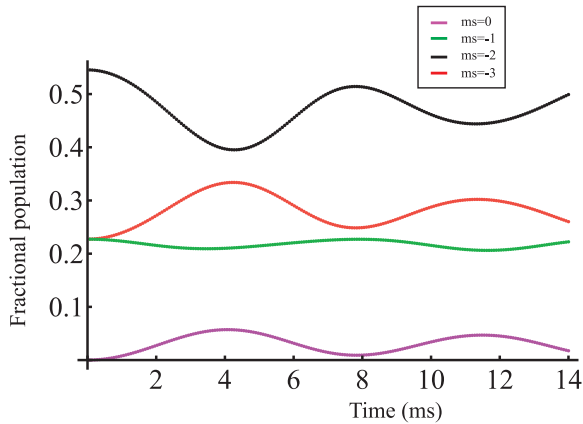}
\caption{\setlength{\baselineskip}{6pt} {\protect\scriptsize \textbf{Results of our toy model for the intersite spin dynamics of doublons.} The initial state is a pure molecular state ($S=6,M=-4$).}} \label{ToyModel}
\end{figure}

In principle, as mentioned above, the coupling constant $V_{I}$ depends on the relative orientation between the two sites and the magnetic field.
An effective value of $V_{I}$ for a two-site model may be obtained from first-order time-dependent perturbation theory.
At short times the initial many-body state is independently coupled to orthogonal many-body states through pairwise interactions.
Matching the time-dependence of the population at a given site with that obtained from the two-site model, we get an effective
$V_I=\sqrt{\sum_{\bf n\neq (0,0)}V_{\bf n}^2}$. For our lattice we obtain $V_I\simeq 2.7 \frac{d^2}{4\pi (\lambda/2)^3}$.
Figure \ref{ToyModel} shows the results of the effective two-site model. Although the qualitative behavior of the populations matches well the dynamics observed in the experiment (see Fig~4 of the article),
the experimental time scale is about 2.5 times faster, showing that the exact dynamics observed in the experiment, although
related to spin-exchange between doubly-occupied sites,  cannot be recovered by a simple two-site model. A more refined many-body theoretical treatment would be needed to match our observations. We note that if one scales the typical timescale of the singlon dynamics calculated within the exact diagonalization method described in S2 (15 ms) to a 4 times faster spin dynamics (due to DDIs between doubly occupied sites), one finds a time scale of 4 ms which matches the observed timescale.

\paragraph{S4.Measurement of the quadratic light shift (QLS) during the spin evolution}
\subparagraph{}
As mentioned in the main text, our system presents a residual QLS, which may significantly handicap inter-site spin-exchange.
This residual QLS may result both from the 532 nm lattice beams and the 1075 nm trapping beam. For these far detuned lasers from any resonances from the ground state, non negligible QLS can arise from small mismatches between the fluorescence rates of states belonging to the same fine structure \cite{EPJD}.

Calculations cannot provide a reliable estimate of the QLS, as they are very sensitive to spectroscopic data. Indeed, calculations based on NIST data \cite{EPJD} predict a QLS of 1 kHz for a 532 nm laser, $\pi$ polarized, with an intensity corresponding to a lattice depth of 25 $E_R$. On the other hand using more recent spectroscopic data \cite{SpectroCr} we obtain for the same laser parameters QLS=$(100\pm400)$ Hz.

\begin{figure}[!h]
\centering
\includegraphics[width=2.5in]{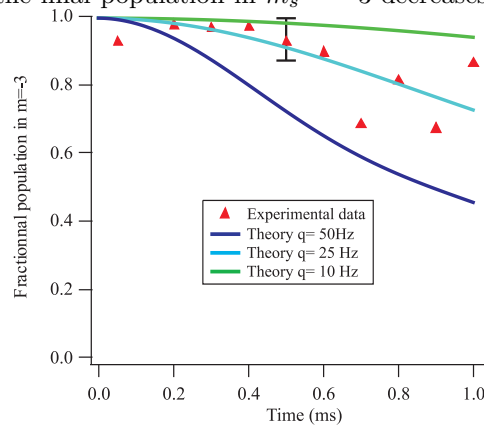}
\caption{\setlength{\baselineskip}{6pt} {\protect\scriptsize \textbf{Interferometric measurement of the residual quadratic light shift.} The final population in $m_s=-3$ is plotted as a function of the duration $T$ of the interferometric sequence. Red traingles: experiment. Full lines: results of simulation for three different QLS.
 }}
\label{Interferometry}
\end{figure}

To measure the residual QLS we apply to the atomic sample polarized in $m_s=-3$ at a magnetic field corresponding to a
Larmor frequency of $30$ kHz,  a symmetric Radio Frequency (RF) interferometric sequence
consisting of a $\pi/2$ pulse, a waiting time $T$, a $\pi$ pulse, a second waiting time $T$, and a final $\pi/2$ pulse.
This spin-echo technique allows to cancel inhomogeneous magnetic field effects. For a small detuning of the RF frequency with respect to the Larmor frequency, the atoms remain in $m_s=-3$ in absence of QLS. In contrast, in
presence of QLS, the final population in $m_s=-3$ decreases to zero for a value of $T$ scaling as the inverse of the QLS.
In Fig.~\ref{Interferometry} we depict the final population in $m_s=-3$ as a function of $T$, comparing the results with our calculations for three different values of the QLS.
From these interferometric results we estimate an upper value for the QLS, $q<25$ Hz.


\vspace{0.5cm}

\begin{thebibliography}{99}

\bibitem{Auerbach} Auerbach, A., Interacting Electrons and Quantum Magnetism. \textit{Springer-Verlag},   N.Y (1994)

\bibitem{RevBloch}Bloch, I., Dalibard, J. and Zwerger, W. Many-body physics with ultracold gases. \textit{Rev. Mod. Phys.} \textbf{80}, 885 (2008)

\bibitem{simulator} Bloch, I., Dalibard, J.,  and Nascimbene, S. Quantum simulations with ultracold quantum gases. \textit{Nature Physics} \textbf{8}, 267 (2012)

\bibitem{Review-Ueda} Y. Kawaguchi and Ueda, M. Spinor Bose–Einstein condensates. Physics Reports {\bf 520}, 253 (2012).

\bibitem{ReviewDSK} Stamper-Kurn, D. and Ueda, M. Spinor Bose gases: Explorations of symmetries, magnetism and quantum
dynamics. ArxiV 1205.1888 (2012)


\bibitem{LatticeDemler} Imambekov, A., Lukin, M. and Demler, E.    Spin-exchange interactions of spin-one bosons in optical lattices: Singlet, nematic, and dimerized phases. \textit{Phys. Rev. A} \textbf{68}, 063602 (2003)

\bibitem{SUN2} Rapp, A., Zarand, G., Honerkamp, C., and Hofstetter, W. Color Superfluidity and "Baryon" Formation in Ultracold Fermions. \textit{Phys. Rev. Lett.} \textbf{98}, 160405 (2007)

\bibitem{SUN1} Hermele, M. , Gurarie, V., and Rey, A. M. Mott Insulators of Ultracold Fermionic Alkaline Earth Atoms: Underconstrained Magnetism and Chiral Spin Liquid. \textit{Phys. Rev. Lett.} \textbf{103}, 135301 (2009)

\bibitem{CrBEC} Griesmaier, A., Werner, J., Hensler, S., Stuhler J. and Pfau, T. Bose-Einstein Condensation of Chromium. \textit{Phys. Rev. Lett.} \textbf{94}, 160401 (2005)

\bibitem{CrBECus} Beaufils, Q. et al. All-optical production of chromium Bose-Einstein condensates. \textit{Phys. Rev. A} \textbf{77}, 061601 (2008)

\bibitem{DysprosiumBEC} Lu, M., Burdick, N. Q., Youn, S. H. and Lev, B.L. Strongly Dipolar Bose-Einstein Condensate of Dysprosium. \textit{Phys. Rev. Lett.} \textbf{107}, 190401 (2011)

\bibitem{DysprosiumFermiSea} Lu, M., Burdick, N.Q. and Lev, B.L. Quantum Degenerate Dipolar Fermi Gas.	\textit{Phys. Rev. Lett.} \textbf{108}, 215301 (2012)

\bibitem{Erbium} Aikawa, K. et al. Bose-Einstein Condensation of Erbium. \textit{Phys. Rev. Lett.} \textbf{108}, 210401 (2012)

\bibitem{Ni08}  Ni, K.K. et al. A High Phase-Space-Density Gas of Polar Molecules. \textit{Science} \textbf{322}, 231 (2008)

\bibitem{Chotia12} Chotia, A., et al. \textit{Phys. Rev. Lett.} \textbf{108}, 080405 (2012).

\bibitem{Wu12} Wu, C.H., Park, J.W., Ahmadi, P., Will, S., and Zwierlein, M.W. Ultracold Fermionic Feshbach Molecules of $^{23}Na^{40}K$. \textit{Phys. Rev. Lett.} \textbf{109}, 085301 (2012)

\bibitem{Heo12} Heo, M.S. et al., Formation of ultracold fermionic NaLi Feshbach molecules.
\textit{Phys. Rev. A} \textbf{86}, 021602(R) (2012)

\bibitem{QMH1} Barnett, R., Petrov, D., Lukin, M., and Demler, E., Quantum Magnetism with Multicomponent Dipolar Molecules in an Optical Lattice. \textit{Phys. Rev. Lett.} \textbf{96}, 190401 (2006)

\bibitem{QMH2} Micheli, A., Brennen, G.K. and Zoller, P. A toolbox for lattice-spin models with polar molecules. \textit{Nature Physics} \textbf{2}, 341 (2006)

\bibitem{QMH3} Gorshkov, A.V. et al. Tunable Superfluidity and Quantum Magnetism with Ultracold Polar Molecules. \textit{Phys. Rev. Lett.} \textbf{107}, 115301 (2011)

\bibitem{QMH4} Peter, D., Muller, S., Wessel, S., and  Buchler, H.P. Anomalous Behavior of Spin Systems with Dipolar Interactions. \textit{Phys. Rev. Lett.} \textbf{109}, 025303 (2012)

\bibitem{Carr} Wall, M.L., Maeda, K., and Carr, L.D. Simulating quantum magnets with symmetric top molecules  arXiv:1305.1236

\bibitem{RevueDipQ3} Baranov, M. A., Dalmonte, M., Pupillo, G. and Zoller, P. Condensed Matter Theory of Dipolar Quantum Gases. \textit{Chemical Reviews} \textbf{112}, 5012 (2012)

\bibitem{Trotzky} S. Trotzky et al., \textit{Science} \textbf{319}, 295 (2008).

\bibitem{BECinterfero} Fattori, M. et al. Magnetic Dipolar Interaction in a Bose-Einstein Condensate Atomic Interferometer. \textit{Phys. Rev. Lett.} \textbf{101}, 190405 (2008)

\bibitem{CrBECStability} Muller S. et al. Stability of a dipolar Bose-Einstein condensate in a one-dimensional lattice. \textit{Phys. Rev. A} \textbf{84}, 053601 (2011)


 \bibitem{demagnetization} Pasquiou B. et al, Spontaneous demagnetization of a dipolar spinor Bose gas at ultra-low magnetic field. \textit{Phys. Rev. Lett.} \textbf{106}, 255303 (2011)

\bibitem{EdH} Kawaguchi, Y., Saito, H., and Ueda, M.  Einstein- De Haas Effect in Dipolar Bose-Einstein Condensates. \textit{Phys. Rev. Lett.} \textbf{96}, 080405 (2006)

\bibitem{Crtube} Pasquiou B. et al, Spin Relaxation and Band Excitation of a Dipolar Bose-Einstein Condensate in 2D Optical Lattices. \textit{Phys. Rev. Lett.} \textbf{106}, 015301 (2011)

\bibitem{resonances} de Paz, A. et al. Resonant demagnetization of a dipolar BEC in a 3D optical lattice. \textit{Phys. Rev. A} \textbf{87},  051609(R) (2013)

\bibitem{Giovanazzi} Hensler, S., et al. Dipolar relaxation in an ultra-cold gas of magnetically trapped chromium atoms. \textit{Applied Physics B} \textbf{77}, 765 (2003)


\bibitem{Kapitza} Gould, P. L., Ruff, G.A., and Pritchard, D.E. Diffraction of atoms by light: The near-resonant Kapitza-Dirac effect. \textit{Phys. Rev. Lett.} \textbf{56},
827 (1986)

\bibitem{Greiner} Greiner, M., Mandel, O., Esslinger, T., Hansch, T.W., and Bloch, I. Quantum phase transition from a superfluid to a Mott insulator in a gas of ultracold atoms. \textit{Nature} \textbf{415}, 39 (2001)

\bibitem{Zwerger02}  Zwerger, W. Mott– Hubbard transition of cold atoms in optical lattices. \textit{J. Opt. B: Quantum Semiclass. Opt.} \textbf{5}, S9 (2003).

\bibitem{weddingcake} DeMarco, B., Lannert, C., Vishveshwara, S., and Wei, T.C. Structure and stability of Mott-insulator shells of bosons trapped in an optical lattice. \textit{Phys. Rev. A} \textbf{71}, 063601 (2005)

\bibitem{Widera} Widera, A., et al. Coherent Collisional Spin Dynamics in Optical Lattices. \textit{Phys. Rev. Lett.} \textbf{95}, 190405 (2005)

\bibitem{Stringari} Pitaevskii, L. and Stringari, S. Bose-Einstein Condensation (Oxford University Press, Oxford) (2003)

\bibitem{PRA RD} Pasquiou B. et al, Control of dipolar relaxation in external fields. \textit{Phys. Rev. A} \textbf{81}, 042716 (2010)

\bibitem{Krauser} Krauser J. S. et al, Coherent multi-flavour spin dynamics in a fermionic quantum gas. \textit{Nature Physics} \textbf{8}, 813–818 (2012)

\bibitem{Krems} Xiang, P., Litinskaya, M., and Krems, R.V. Tunable exciton interactions in optical lattices with polar molecules. \textit{Phys. Rev. A} \textbf{85}, 061401(R) (2012)


\bibitem{Ye} Yan, B., et al. Realizing a lattice spin model with polar molecules. arXiv:1305.5598


\end{thebibliography}

\begin{thebibliography}{99}
\bibitem{Gajda} Swislocki, T. et al. Tunable dipolar resonances and Einstein-de Haas effect in a $^{87}$Rb-atom condensate. \textit{Phys. Rev. A} \textbf{83}, 063617 (2011)
\bibitem{EPJD} Chicireanu, R. et al. Accumulation of chromium metastable atoms into an Optical Trap. \textit{EPJD} \textbf{45}, 189 (2007)
\bibitem{SpectroCr} Sobeck J. S., Lawler J.E. and Sneden, C. Improved laboratory transition probabilities for neutral chromium and determination of the chromium abundance for the sun and three stars. The Astrophysical Journal \textbf{667}, 1267 (2007)

\end{thebibliography}
\end{document}